\documentclass[twocolumn,a4paper,showpacs,preprintnumbers,aip,jap,amssymb,amsmath,showkeys,10pt]{revtex4-1}

\usepackage[francais,english]{babel}
\usepackage[utf8]{inputenc}
\usepackage[T1]{fontenc}
\usepackage{fouriernc}

\usepackage{amsmath}
\usepackage{textcomp}
\usepackage{natbib}
\usepackage{graphicx}
\usepackage{dcolumn}
\usepackage{bm}

\begin{document}

\title[18pt]{Impact of current paths on measurement of tunneling magnetoresistance and spin torque critical current densities in GaMnAs-based magnetic tunnel junctions}

\author{A. Ben Hamida}
\affiliation{\rm{Physikalisch-Technische Bundesanstalt}, \emph{Bundesallee 100, D-38116 Braunschweig, Germany}}
\author{F. Bergmann}
\affiliation{\rm{Physikalisch-Technische Bundesanstalt}, \emph{Bundesallee 100, D-38116 Braunschweig, Germany}}
\author{K. Pierz}
\affiliation{\rm{Physikalisch-Technische Bundesanstalt}, \emph{Bundesallee 100, D-38116 Braunschweig, Germany}}
\author{H.W. Schumacher}
\affiliation{\rm{Physikalisch-Technische Bundesanstalt}, \emph{Bundesallee 100, D-38116 Braunschweig, Germany}}

\begin{abstract}
GaMnAs-based magnetic tunnel junction (MTJ) devices are characterized by in-plane and perpendicular-to-plane magnetotransport at low temperatures. Perpendicular-to-plane transport reveals the typical tunneling magnetotransport (TMR) signal. Interestingly, a similar TMR signature is observed in the in-plane transport signal. Here, low-ohmic shunting of the MTJ by the top contact results in significant perpendicular-to-plane current paths. This effect allows the determination of TMR ratios of MTJs based on a simplified in-plane measurement. However, the same effect can lead to an inaccurate determination of resistance area products and spin torque critical current densities from perpendicular-to-plane magnetotransport experiments on MTJ pillar structures.   
\end{abstract}

\maketitle

\section{Introduction}

Metallic magnetic tunnel junctions (MTJs) are important building blocks of spintronic devices such as magnetic random access memories or hard disk drive read heads. Also MTJs based on ferromagnetic semiconductors could be promising for future device applications as they could combine magnetic memory and semiconductor logic functions in the same class of materials. Presently, GaMnAs can be considered the most prominent prototype of diluted magnetic semiconductors \cite{Jungwirth_RevModPhys78_809_2006}. A variety of GaMnAs-based MTJs with various barrier materials have been studied so far: AlAs \cite{Tanaka_PRL87_026602_2001, Ohya_PSSC3_4184_2006}, GaAs \cite{Chiba_PhysicaE21_966_2004}, InGaAs \cite{Elsen_PRB73_035303_2006}, ZnSe \cite{Saito_PRL95_086604_2005} and AlMnAs \cite{Ohya_APL95_242503_2009}. The electronic key parameters such as the tunneling magnetoresistance (TMR) ratio of such semiconducting MTJs are usually determined by perpendicular-to-plane magnetoresistance (PPMR) measurements between the top and the bottom contact of a micropatterned MTJ pillar. However, during such PPMR measurements electrical contacting of the top contact, e.g., by wire bonding, could lead to an electrostatically or mechanically induced modification or even destruction of the MTJ's the delicate tunnel barrier yielding wrong TMR parameters.\\ 

Here we show that low-ohmic shunting of the in-plane MTJ resistance by a metallic top contact can induce significant perpendicular-to-plane current components. This effect can be used to derive the TMR from in-plane magnetoresistance (IPMR) measurements of large MTJ pillar structures thereby avoiding the aforementioned problems. Based on finite element simulations we further show that such shunting can also induce strongly inhomogeneous current densities in PPMR measurements of micron or sub-micron scale MTJ pillars. This effect has to be taken into account for the reliable determination of the resistance area product (RA) and the spin torque \cite{Slonczewski_JMMM159_L1_1996, Berger_PRB54_9353_1996, Myers_Science285_867_1999, Katine_PRL84_3149_2000} critical current density $j_c$ from PPMR measurements.

\section{Samples}

Our GaMnAs MTJ stacks are grown in a low-temperature MBE environment on a semi-insulating GaAs wafer. Details of our GaMnAs growth procedure are published elsewhere \cite{BenHamida_JMSJ36_49_2012}. The stack \mbox{sequence} consists of 100$\ $nm carbon-doped $p^{+}$-GaAs as a non-magnetic bottom contact line and a junction of \mbox{GaMnAs}(100$\ $nm)/barrier/GaMnAs(50$\ $nm) with about 5\% of Mn concentration in both GaMnAs layers. The barrier of the MTJ stack consists of $2\,\textrm{nm}$ thick AlAs sandwiched between two $1\,\textrm{nm}$ thick layers of GaAs. The wafers are fabricated into rectangular pillars with $300\,\mu{}\textrm{m}$ x $300\,\mu{}\textrm{m}$ lateral extensions as shown in the microscope image of Fig. 1(a). The samples consist of an MTJ pillar with one top contact (TC), two bottom contacts (BC) and a GaAs:C bottom lead between the two BCs and the bottom MTJ layer. A sectional sketch of the sample is shown in Fig. 1(b). \\

The fabrication of the pillar structures is carried out by multiple e-beam lithography steps in combination with lift-off and wet etching. The ohmic TC and BC consist of $10\,\textrm{nm}$ Ti and $100\,\textrm{nm}$ Au. The TC is directly deposited on the uppermost GaMnAs layer of the MTJ whereas the BC contacts the highly C-doped GaAs lead via a remaining 30$\ $nm thick GaMnAs layer. This allows the realization of well defined ohmic contacts to the GaAs:C. For magnetotransport measurements the samples were wire-bonded and placed in a commercial He cryostat with variable temperature insert allowing the application of 3D magnetic vector fields up to $1\,\textrm{T}$.
\\

\begin{figure}
\centering\includegraphics[width=\linewidth]{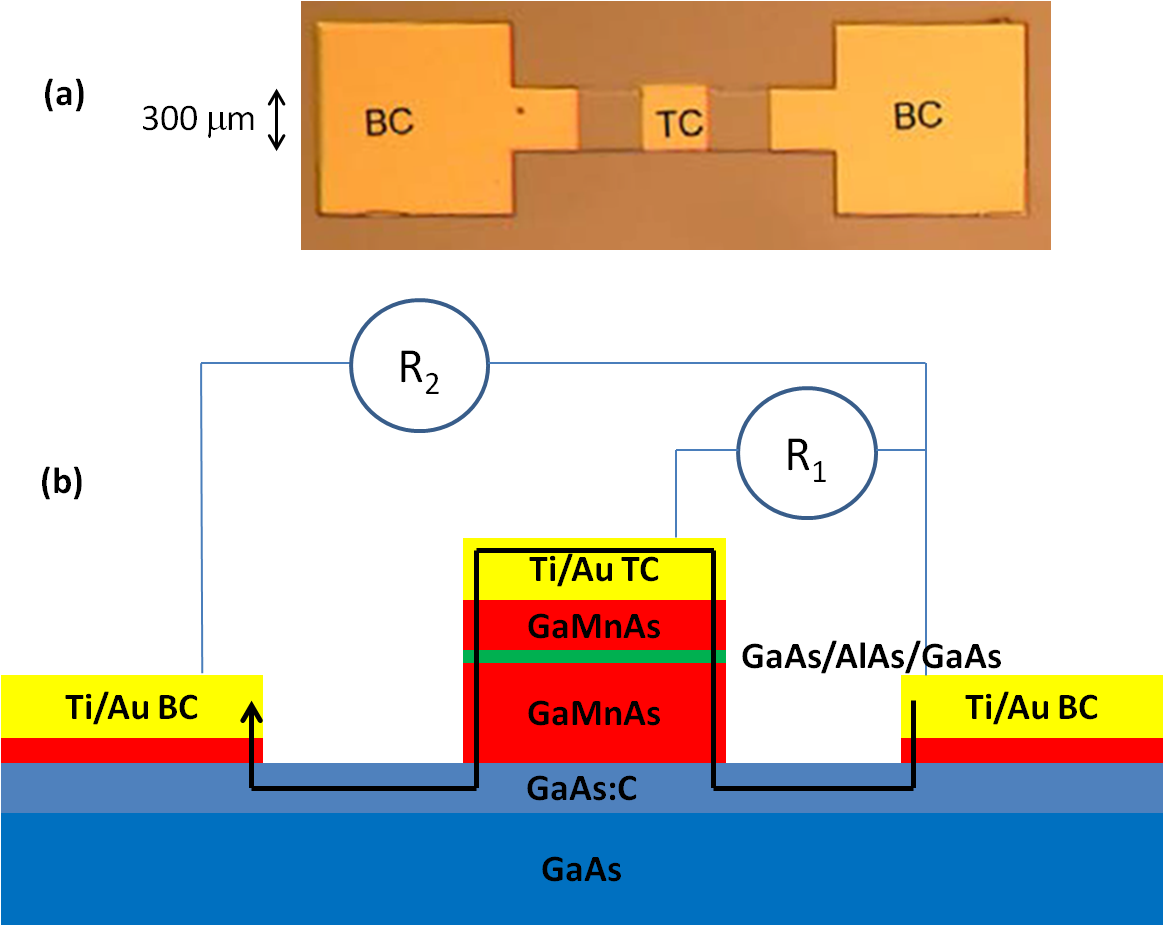}
\renewcommand{\figurename}{Fig.}
\caption{(a) Optical image of the patterned MTJ pillar structure. (b) Sectional sketch of MTJ pillar structure with $p^{+}$-GaAs:C bottom contact line and gold contacts to the contact line and top MTJ layer. Electric measurement configurations for IPMR measurements between outer BCs ($R_2$) and PPMR measurements between one BC and TC ($R_1$) are sketched. Black arrow indicates the suggested current path for measurement of $R_2$ due to shunting of the MTJ by the top gold contact.}
\end{figure}

\section{Experimental results}

All magnetotransport measurements were performed at $T = 10\,\textrm{K}$. For PPMR measurement of ($R_1$), a current of $I = 100\,\mu{}A$ was applied between the TC of the pillar and one of the BCs \cite{Ohno_PRL68_2664_1992, Ohno_APL69_363_1996}. As sketched in Fig.$\ $1(b), $R_1$ contains a contribution of the pillar and a contribution of the highly C-doped GaAs back contact lead. For a complete characterization of the hysteresis and anisotropies of the MTJ pillar magnetoresistance loops were taken along various in-plane angles to \mbox{characterize} the field dependence along the easy and hard anisotropy axes. For in-plane easy axis loops, the field is applied along the [001] crystalline orientation. Fig.$\ $2(a) shows a corresponding easy axis loop with the magnetic field varied between $-30\,\textrm{mT}$ and $+30\,\textrm{mT}$. The resistance $R_1$ shows three magnetization states corresponding to three distinct resistance states: I) parallel configuration $R_{1,p}$ (low resistance state), II) 90\degres{} configuration (intermediate resistance state) and III) anti-parallel configuration (high resistance state). The corresponding states are sketched by the arrows in Fig.$\ $2. A TMR ratio $\frac{\Delta{R_1}}{R_{1,p}} \approx$ 1\% is extracted from the resistance difference $\Delta{}R_1 = 7\,\Omega$ between the parallel and the anti-parallel configuration. Note that this TMR value also contains the resistance contributions from the contacts and the GaAs:C bottom leads.\\

\begin{figure}
\centering\includegraphics[width=\linewidth]{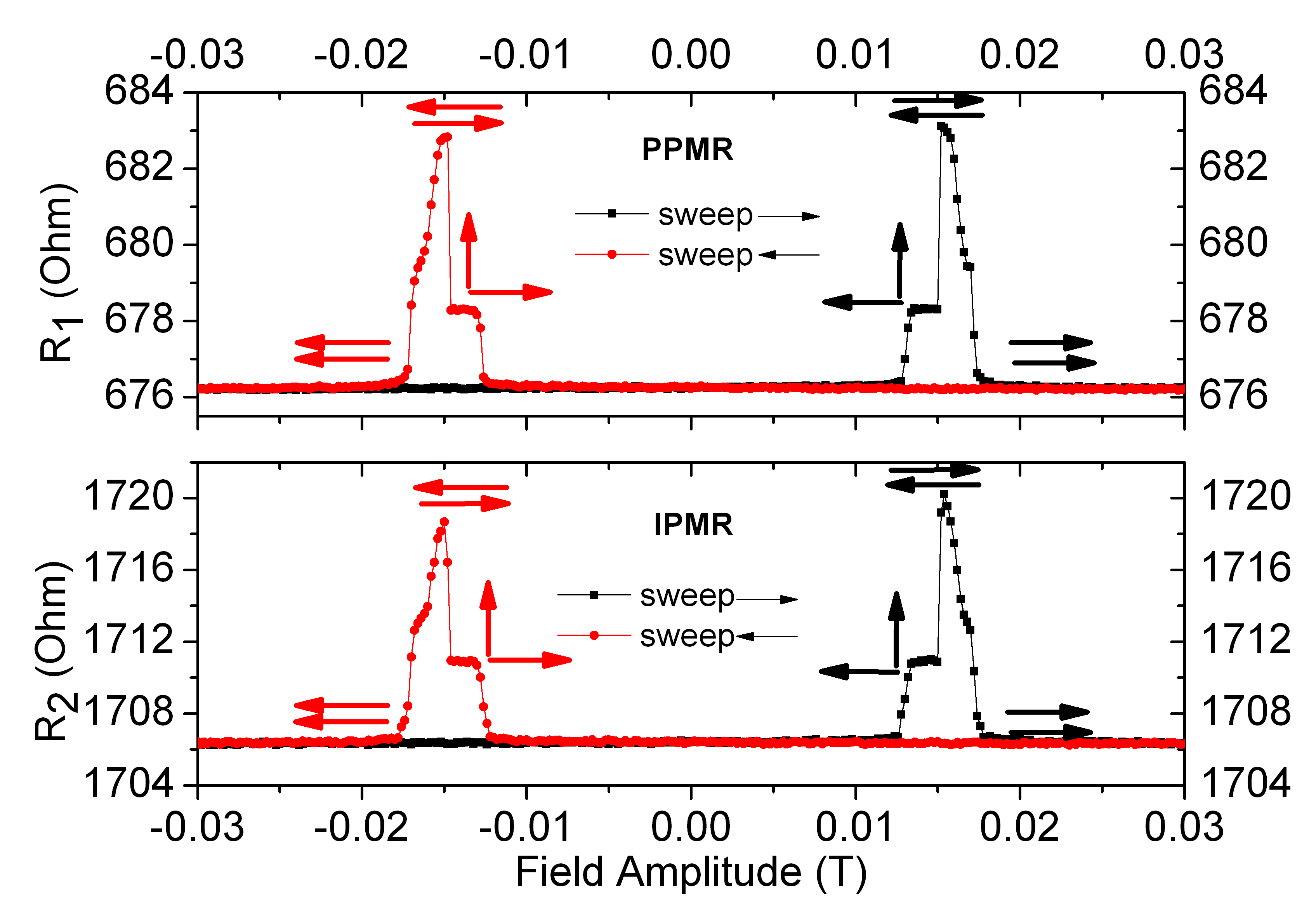}
\renewcommand{\figurename}{Fig.}
\caption{Magnetoresistance hysteresis loop for a $300\,\mu{}\textrm{m}$ x $300\,\mu{}\textrm{m}$ pillar at $T = 10\,\textrm{K}$ and at fixed in-plane field angle along [001] (easy-axis field). (a) PPMR measurement of $R_1$. (b) IPMR measurement of $R_2$. Black curves: field swept from negative to positive fields; red curves: field swept from positive to negative fields.}
\end{figure}

Figure 2(b) shows a measurement of $R_2$ of the same device. As sketched in Fig. 1, $R_2$ is the IPMR measured between the two BCs to the C-doped GaAs bottom lead of the TMJ pillar. Here, no significant in-plane \mbox{magnetoresistance} can be naively expected. However, surprisingly the measurement of $R_2$ shows a very similar behavior as $R_1$ with a similar TMR ratio but twice the resistance difference: $\Delta{}R_2 = 14\,\Omega \approx{2\,\Delta{}R_1}$. This similarity suggests a current path which leads twice through the TMR structure as sketched by the black arrow in Fig. 1(b). The low-ohmic gold TC of the pillar has a resistivity which is about two orders of magnitude lower than the resistivity of the bottom GaAs:C layer below the pillar. Therefore, the Au TC could shunt the resistance of the bottom lead and MTJ stack. For this current path the TMR of the MTJ is measured twice in series explaining the factor of two between both resistance differences, $\Delta{}R_1$ and $\Delta{}R_2$ respectively, in a straightforward manner. To confirm this scenario similar MTJ pillars without the central Au TC were fabricated and characterized (not shown). These samples do not show a significant TMR signal in the IPMR measurements. This is expected as no \mbox{shunting} by a low ohmic TC is present in these devices. In the above measurements of the shunted devices basically the same TMR ratio is derived for both IPMR and PPMR \mbox{measurements}. This could allow an alternative way to measure the TMR of a MTJ in a planar geometry.  The practical advantage of such a configuration is the absence of a bonding wire on the pillar TC. This would reduce the risk of damaging the delicate tunneling structures during the bonding process and migth allow simplified device fabrication.

\section{Simulations and discussions}

The above low-ohmic shunting should result in a strongly inhomogeneous current density near the pillar edges. To confirm this and to estimate the length scale of this inhomogeneous current distribution two dimensional finite element simulations of the current distribution were carried out using a commercial simulation package \cite{COMSOL_Website}. In the simulations of the current density distribution a two dimensional section of the device with a $20\,\mu{}m$ pillar was simulated. The resulting current distribution was calculated upon voltage application between the two BCs. The simulation parameters of the structure were based on the experimentally derived conductivity parameters of the semiconductor layers and on literature values of the Au contacts. The effective conductivity of the AlAs barrier, $G$, was not accessible experimentally. Therefore the tunnel barrier was \mbox{modeled} as having an anisotropic conductivity with vanishing conductance in-plane (x-direction) and non-vanishing conductance $G$ perpendicular-to-plane (y-direction). $G$ was varied from 0.1 to 100 S/m and the resistance of the complete device was calculated. From the comparison of the calculated resistance to the experimental values we can estimate that that $G$ of our devices is between 1 and 10 S/m. \\ 

Figure 3 shows typical maps of the simulated current density $j$ distribution during such IPMR measurements. In the figure the current distribution near the edge of the MTJ pillar is shown. The simulated results are shown for four different values of the tunneling barrier conductance of $G =$ 0.1 S/m, 1 S/m, 10 S/m, and 100 S/m (from top to bottom). The local orientation of the current density $j$ is indicated by the white arrows. The amplitude of $j$ is logarithmically encoded by the length and thickness of the arrows. The color of the map encodes $j_y$, the y-component of $j$. Blue represents $j_y = 0$ whereas red \mbox{represents} a positive high value, i.e., current flowing upwards. \\

\begin{figure}
\centering\includegraphics[width=\linewidth]{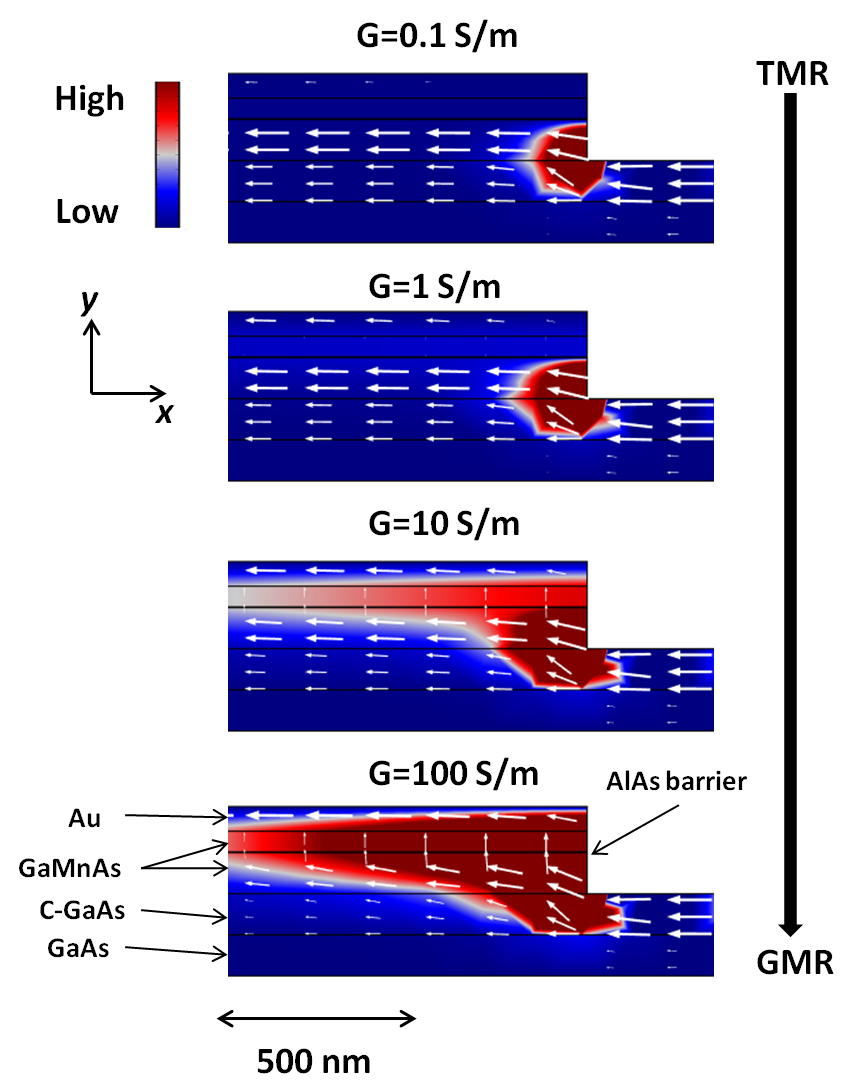}
\renewcommand{\figurename}{Fig.}
\caption{Simulated local current density $j$ during IPMR ($R_2$) measurements near the edge of the $20\,\mu{}\textrm{m}$ MTJ pillar. Results are shown for four different effective conductivities $G$ of the tunnel barrier. Arrows indicate the local orientation of $j$. The arrow dimensions scale logarithmically with the absolute value of $j$. Color encodes $j_y$. Blue: $j_y = 0$; red: $j_y > 0$; current flows from bottom to top.}
\end{figure}

From Fig.$\ $3, it is obvious that the value of $G$ plays an important role for the above shunting scenario. The four simulations cover the range from high resistance TMR-like stacks (top) to low resistance GMR-like stacks (bottom) as indicated in the figure. In all four cases $j_y$ is locally increased near the pillar edges. Here, the current is injected from the thin GaAs:C lead into the thicker MTJ and a part of the current flows upwards (red region on right hand side). However, for the uppermost panel with $G =$ 0.1 S/m only very little current passes the tunnel barrier. The in-plane current is mainly concentrated in the lower GaMnAs layer of the MTJ. Also a significant portion of the current flows in the GaAs:C. In contrast, practically no current flows in the TC gold layer, at least in the displayed part of the MTJ pillar. For this tunnel conductance, the shunting effect should thus be small or negligible. Also for the case of $G =$ 1 S/m most of the current still flows in the lower GaMnAs layer and in the GaAs:C. However now a small part of the in-plane current starts flowing in the TC (small white arrows), especially with increasing distance from the pillar edge. For $G =$ 10 S/m the current distribution is significantly different. Now the tunnel barrier resistance is low enough to allow shunting by the TC. Near the edge of the MTJ a significant part of the current passes the tunneling barrier leading to a locally increased current density through the barrier (red region). Also a significant part of the in-plane current flows in the TC. For such parameters, a significant TMR contribution should be present in the IPMR measurements as found experimentally. Finally, for $G =$ 100 S/m this effect is more pronounced. Near the pillar edge most of the current from the bottom lead flows perpendicularly through the MTJ to the TC. The dominant part of the in-plane current is concentrated in the TC and a complete shunting occurs already on a short length scale below $1\,\mu{}m$.\\

Such strong shunting effect will also influence the current distribution in PPMR measurements of large MTJ pillars. For such measurements, a strongly inhomogeneous current distribution should occur. Figure 4 shows results of simulations of the current density $j$ for PPMR measurements of an MTJ pillar with $20\,\mu{}m$ diameter. Figure 4(a) shows a color map of the current density distribution near the MTJ pillar edge for $G =$ 10 S/m. The color and arrow representation is the same as in Fig.$\ $3. Also for this PPMR simulation the perpendicular current $j_y$ through the tunneling barrier is significantly enhanced near the edge of the pillar from where the current is injected (red region). The resulting current density is thus strongly inhomogeneous. \\

\begin{figure}
\centering\includegraphics[width=\linewidth]{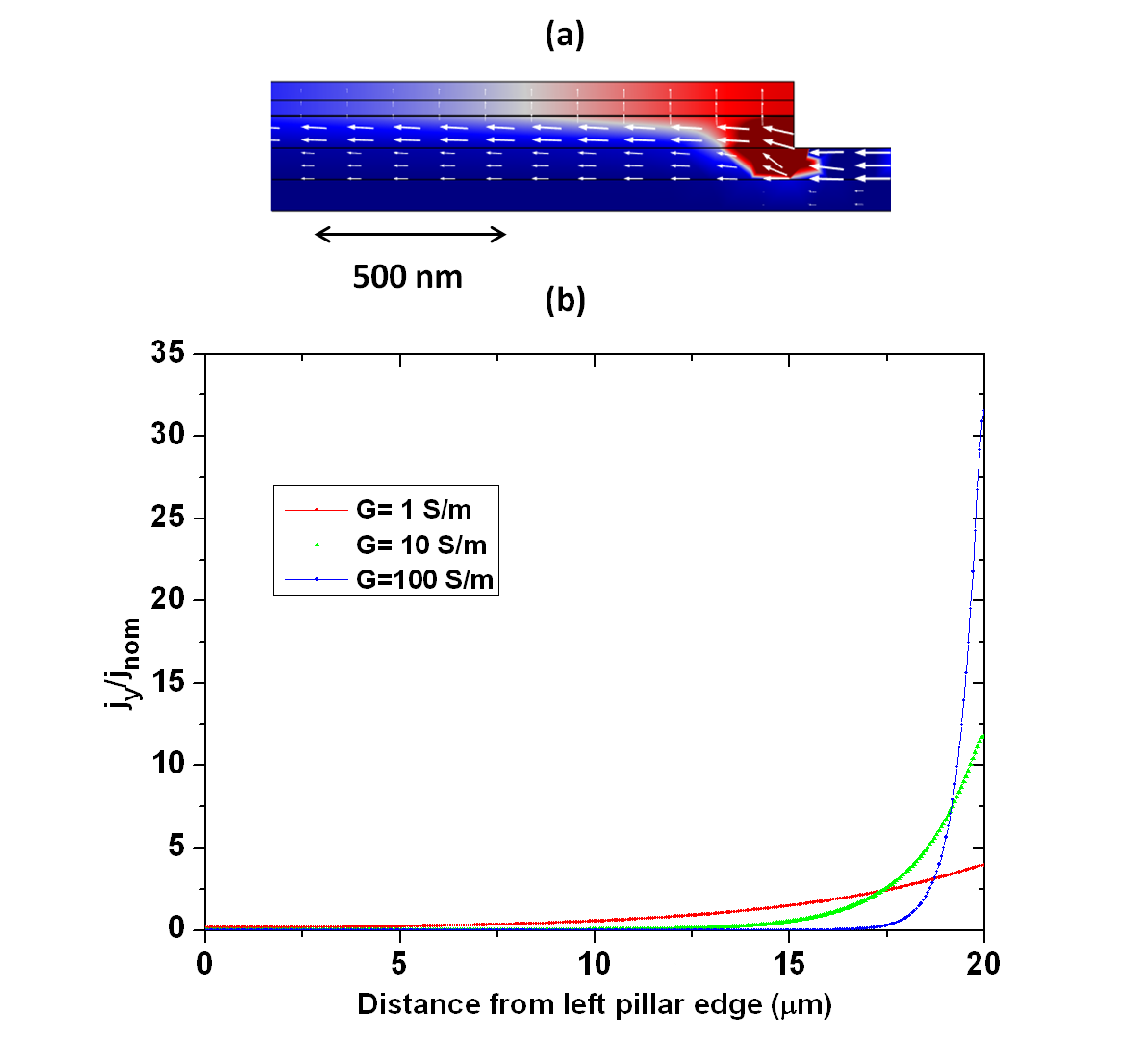}
\renewcommand{\figurename}{Fig.}
\caption{Simulation of the current density in PPMR measurements of a 
$20\,\mu{}m$ MTJ pillar. (a) Color map of the current density $j$ for a barrier conductivity of $G=10\,S/m$. Same color and arrow representation as in Fig.$\,$3. (b) Normalized y-component of $j_y$ as function of the lateral distance (x-direction) from the pillar edge for $G = 1, 10,$ and $100$. $j_y$ is strongly enhanced near the pillar edge.}
\end{figure}

The length scale of the inhomogeneity again depends on the value of $G$ as shown in Fig. 4(b). Here the \mbox{normalized} $j_y$ at the position of the tunnel barrier is plotted as function of the lateral distance from the right pillar edge. $j_y$ is normalized by the nominal average current density $j_{nom} = \frac{I}{A}$ assuming homogeneous current distribution. Here $I$ is the total current through the device and $A$ is the junction area. For an ideal homogeneous current distribution the normalized $j_y$ should have a constant value of $1$ independent of the lateral position. However for the three given simulation parameters of $G =$ 1, 10, and 100 S/m the current density distribution significantly differs from this idealized situation. In all cases $j_y$ is significantly enhanced near the pillar edge. For $G=$ 1 S/m a maximum current density of $\frac{j_y}{j_{nom}} \approx 4$ occurs at the pillar edge. The current density decays over a distance of about 10 $\mu{}m$. With increasing $G$ the increase of $j_y$ becomes more pronounced and more localized near the edge. For $G =$ 10 S/m, $\frac{j_y}{j_{nom}}$ reaches a maximum of about 12 and decays on a scale of about 5 $\mu{}m$. For $G =$ 100 S/m, $\frac{j_y}{j_{nom}}$ locally increases up to about 32 and decays on a on a scale of only about $2\,\mu{}m$. Especially for low resistance MTJ systems such strongly inhomogeneous current density must be taken into consideration when determining the RA product from PPMR measurements. The effective area $A^*$ through which the current flows is significantly smaller than the lithographic area $A$ of the MTJ sample. Hence the RA product derived from such measurement under the assumptions of a homogeneous current distribution would yield too large values.\\

For measurements of the critical current density $j_0$ for spin torque magnetization reversal \cite{Chiba_PRL93_216602_2004, Wunderlich_PRB76_054424_2007, Mark_PRL106_057204_2011} this inhomogeneous current distribution could play an important role. Here $j_c$ is generally determined from PPMR measurements on micron or sub-micron sized MTJ pillars. $j_c$ is then derived by measuring the so-called critical switching current $I_c$ which induces spin torque magnetization reversal. Under the assumption of a constant homogeneous current distribution ($j_{nom} =$ const) $j_c$ is given by $j_c = \frac{I_c}{A}$. However, even for small junction diameters the simplified assumption of a homogeneous current distribution is not necessarily valid.\\

Figure 5 shows a simulation of the current distribution of a PPMR measurement for an MTJ pillar of 1 $\mu{}m$ \mbox{diameter}. The diameter is thus of the same order as in the experimental measurements of $j_c$ in GaMnAs-based MTJs \cite{Chiba_PRL93_216602_2004, Elsen_PRB73_035303_2006, Watanabe_APL92_082506_2008}. Again Fig. 5(a) shows a color map of $j_y$ whereas (b) shows the normalized $j_y$ over the lateral extension of the pillar. In (a) the current density map of the MTJ pillar during the PPMR measurement is shown for $G =$ 10 S/m. Also for this small pillar structure a \mbox{significant} enhancement of the $j_y$ near the pillar edge on the injection side is present.\\ 

Figure 5(b) allows quantifying this effect for different values of $G$. For a small $G=$ 1 S/m (TMR-like device) only a weak effect is present. Here $j_y$ deviates by less than 1\% from the nominal value $j_{nom} = \frac{I}{A}$. A PPMR spin torque switching experiment on such a structure would thus lead to a correct determination of $j_c$. Also for $G =$ 10 S/m the assumption of a constant $j_y = j_{nom}$ is feasible. Here $j_y$ only deviates by about $\pm{10 \%}$ from the nominal value. However for a low resistive (strongly GMR-like) junction this assumption is no longer valid. Here the \mbox{local} $j_y$ is increased by up to 70\% with respect to the nominal value near the injection edge. In contrast on the other side of the pillar a decrease by about 40\% is found. \mbox{During} a PPMR spin torque measurement on such sample $j_y$ would first exceed $j_c$ near the pillar edge. This could lead to the local nucleation of a reversed domain wall by spin torque reversal and a subsequent full reversal by domain wall propagation. Such mechanism could result in an underestimation of $j_c$ from the PPMR measurements by several tens of per cent.

\begin{figure}[!ht]
\centering\includegraphics[width=\linewidth]{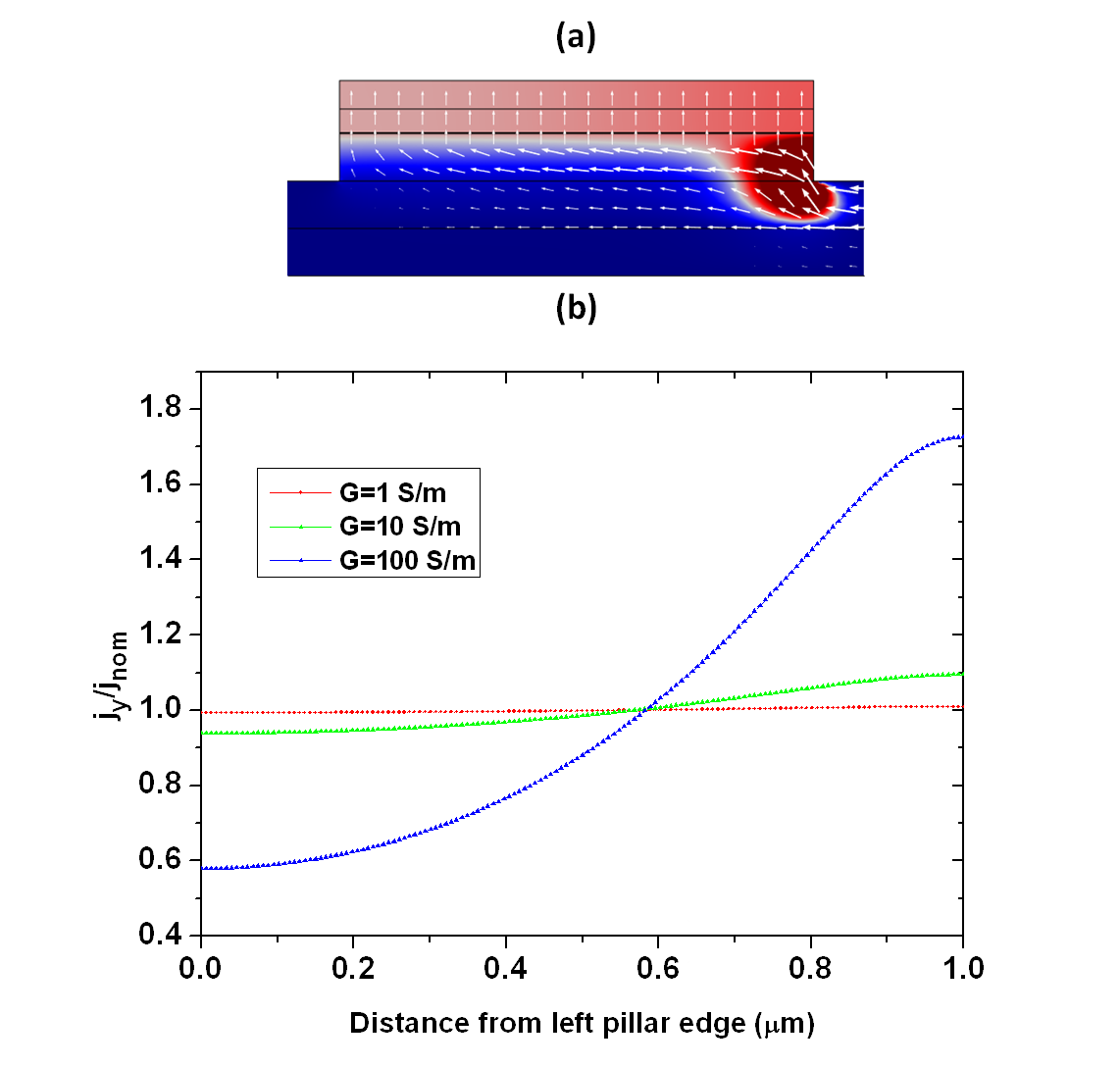}
\renewcommand{\figurename}{Fig.}
\caption{Simulation of the current density in PPMR measurements of a 1 $\mu{}m$ diameter MTJ pillar. (a) Color map of the current density $j_y$ for a barrier conductivity of $G=$10 S/m. Same color and arrow representation as in Fig. 3 and 4. (b) Normalized $j_y$ as function of the lateral distance (x-direction) from the right pillar edge for $G =$ 1, 10, and 100. For $G =$ 100 S/m, $j_y$ is strongly enhanced near the pillar edge.}
\end{figure}

\section{Conclusion}

In conclusion, we have shown that low-ohmic shunting by a metallic top contact can lead to strongly inhomogeneous current paths in transport measurements on \mbox{GaMnAs}-based MTJs. The shunting can induce a strongly enhanced current density located near the pillar edges. This effect is more pronounced for low-ohmic tunnel barriers and thus for GMR-like devices. For low-ohmic tunnel junctions this effect can persist down to very small pillar diameters of 1 $\mu{}m$ diameter and below as relevant for spin torque magnetization reversal experiments. For such devices, the determination of the spin torque critical current density $j_c$ from standard transport measurements could lead to an underestimation of $j_c$ by more than 50\%. This effect should be taken into account for the measurement of RA and for the analysis of spin torque switching experiments for samples showing a significantly higher resistivity of the MTJ stack components than the top contact. However, the effect of this current inhomogeneity could be reduced by using an adapted sample design. Here, e.g., for a circular MTJ pillar with a circular outer bottom contact electrode, the current inhomogeneity should be less significant compared to a linear sample design.

\section{Acknowledgments}

The work was supported by DFG SPP \mbox{Semiconductor} Spintronics and EMRP JRP MetMags. The EMRP is jointly funded by the EMRP participating countries within EURAMET and the EU.
 

\end{document}